\documentclass[prodmode,acmtecs]{acmsmall} % Aptara syntax

\usepackage{url}
\usepackage{natbib}

\setcopyright{acmlicensed}

\issn{2157-6904/2016}

\acmVolume{00}
\acmNumber{00}
\acmArticle{0000}
\acmYear{2016}
\acmMonth{0000}

\doi{http://dx.doi.org/10.1145/2899004}

\begin{document}

% Page heads
\markboth{G. Widmer}{The Con Espressione Manifesto}

% Title portion
\title{Getting Closer to the Essence of Music: \\ The \textit{Con Espressione} Manifesto}

\author{GERHARD WIDMER
 \affil{Johannes Kepler University Linz, Austria }
 % \\ and \\
 %        Austrian Research Institute for Artificial Intelligence (OFAI), Vienna}
}

\begin{abstract}
This text offers a personal and very subjective view on the current situation 
of Music Information Research (MIR). Motivated by the desire to build systems
with a somewhat deeper understanding of music than the ones we currently have,
I try to sketch a number of challenges for the next decade of MIR research,
grouped around six simple truths about music that are probably generally agreed on,
but often ignored in everyday research. 
\end{abstract}




\terms{AI, Machine Learning, Music}


\acmformat{Gerhard Widmer, 2016. Getting Closer to the Essence of Music: The Con Espressione Manifesto.}
% At a minimum you need to supply the author names, year and a title.
% IMPORTANT:
% Full first names whenever they are known, surname last, followed by a period.
% In the case of two authors, 'and' is placed between them.
% In the case of three or more authors, the serial comma is used, that is, all author names
% except the last one but including the penultimate author's name are followed by a comma,
% and then 'and' is placed before the final author's name.
% If only first and middle initials are known, then each initial
% is followed by a period and they are separated by a space.
% The remaining information (journal title, volume, article number, date, etc.) is 'auto-generated'.

\begin{bottomstuff}
% The author gratefully acknowledges generous long-term financial support for this research
% by the Austrian Science Fund FWF (Wittgenstein Award 2009, project number Z159)
% and the European Research Council (ERC Advanced Grant, project number
% ERC-2014-AdG 670035).

Author's addresses: G. Widmer, Dept. of Computational Perception,
Johannes Kepler University Linz, Austria; and Austrian Research
Institute for Artificial Intelligence (OFAI), Vienna, Austria.
{\tt gerhard.widmer@jku.at}
\end{bottomstuff}

\maketitle

\section{About this Manifesto}
\label{sec:about}

This text wishes to point our research community to a set of open problems
and challenges in Music Information Research (MIR) \cite{mires:roadmap},\footnote{
The audience addressed here is all researchers whose
goal is to develop computer systems that deal and interact with musical
contents, or musicians, in ways that are musically meaningful and useful.
That includes fields like Music Information % Retrieval/
Research (MIR),
Sound and Music Computing (SMC),
but also parts of neighbouring fields like musical informatics, computational
music theory, performance research, and even music cognition and psychology.
For notational convenience, I will keep referring to my target field
as MIR, with the implicit understanding that it really encompasses
a broader set of research communities.}
and to initiate new research that will hopefully lead to a
qualitative leap in musically intelligent systems. As a \textit{manifesto},
it presents the author's personal views\footnote{
To emphasise the highly subjective character of this discussion, I have chosen to write the article
in the first person, rather than using the generic ``we'' or the passive voice.}
on the strengths and limitations of current MIR research, on what is missing,
and on what some of the next research steps could be.
The discussion is based on the conviction that musically gratifying
interaction with computers at a high level of musical quality -- which I take
to be the ultimate goal of our research community -- will only be possible if
computers (and, in the process, we) achieve a much deeper understanding
of the very \textit{essence} of music, namely, how music is perceived by, and affects,
human listeners.
This is also the personal manifesto of a music lover who is somewhat dissatisfied
with the level of musical sophistication exhibited by current MIR systems.

The specific motivation for publishing the manifesto at this point in time
is the start of a large research project called \textit{Con Espressione}
(\url{www.cp.jku.at/research/ConEspressione}), generously supported by the European Research
Council (ERC), which hopes to address some of the challenges,
with a particular focus on the \textit{expressive} aspects of music.
But the problems discussed here go way beyond what a single
research project can tackle. Thus, this is a call to the entire MIR community
to consider some of the issues in designing their research agendas.

\section{The Need for a Deeper Understanding of Music in Computers}
\label{sec:need}

The field of MIR (and related fields) has made a lot of progress, has achieved
some spectacular results, and has produced -- and keeps producing -- highly useful
applications in the commercial world of digital music. The contributions in this
volume are evidence of this.
Computers can detect music in complex audio,
% \cite{schlueter:sonnleitner:dafx2012},
can identify, track, classify, and tag songs;
they can extract many structural elements from music signals, such as
onsets, beats, rhythm patterns,
% \cite{krebs:etal:ismir2014},
metrical structure,
% \cite{krebs:etal:ieee2015},
melodic lines, harmonies;
they can segment musical pieces based on sound similarity, local changes, or repetition,
% \cite{paulus:etal:ismir2010,mueller:book2015},
and use this for music summarisation and many other useful services.
A good overview of the technical state of the art can be found in \cite{mueller:book2015}.

Here are a few things our computers can \textit{not} (yet) do:
\begin{enumerate}
 \item Distinguish between songs that I might find boring or interesting.\footnote{
       Of course -- as for all the other examples I am presenting here --
       there is no `true' answer.
       There will be differences in judgment between listeners,
       depending on musical preferences, experience, mood, historical period,
       even social context.
       However, there are fundamental information-theoretic principles at work in
       (aesthetic) perception \cite{meyer:1956,moles:1966}.
       A song consisting of one single chord repeated for three minutes, or a
       melody stepping up and down the major scale ten times in a row,
       will be perceived as monotonous, if not boring, by almost anyone.
       (Note that this is not an aesthetic judgment: monotony
       may be intended and understood as an element of style, even as
       an artistic or political statement;
       think of genres like Minimal Music, Punk Rock, or \textit{`Krautrock'}).}
 \item Return (among other pieces) Beethoven's piano sonata op.81a (\textit{Les Adieux})
       when asked for a piece of classical music with a surprise at the beginning.
 \item Classify Tom Jobim's or Jo\~ao Gilberto's rendition of
       \textit{Garota de Ipanema} as more relaxed and `flowing' than
       Frank Sinatra's.\footnote{... or am I the only one to think so?}
 \item Play along with human musicians (e.g., accompany a soloist in a
       piano concerto) in a musically sympathetic way, recognising, anticipating,
       and adapting to the musicians' expressive way of playing (dramatic, lyrical,
       sober, ...).
\end{enumerate}

For item (1), the computer would have to have an idea of redundancy vs.~unpredictability,
structure vs.~randomness, and the role of expectation in human listening;
for (2), it would have to have (learned) a model of classical music style (in this
particular case, harmony);
for (3), it would have to analyse subtle aspects of performance (timing, intonation,
articulation, etc.), and understand how they contribute to the musical and expressive
character of a performance;
and for (4), it would have to have all of the above capabilities (plus the ability to
recognise and decide in real time).
None of these aspects, I claim, has yet received sufficient attention
in the MIR community, and all of these (and more) would be needed to bring musical
computers closer to understanding the essence of music: how
it affects human listeners.

One of the big, overarching goals for the next decade would thus be to
\textit{equip computers with a deeper `understanding'
of music, its qualities, and how
they are perceived by humans, in order to support a new generation of music
systems and services at a new level of quality}.\footnote{
Of course, producing human-level
performance on some task in a machine does not necessarily require mimicking humans
-- think of chess programs, for instance.
(Thanks to Fran\c{c}ois Pachet for reminding me of this.)
Then again, ``deeper understanding of music'' does not necessarily
mean ``human-like''. What I mean is computational models that
can identify or predict some of the same kinds of patterns in music that
educated human listeners would perceive. Developing such models will require us, as
researchers, to understand more about human music perception.
Or to put it in \cite{herrera:etal:ismir2009}'s words,
``[w]e will only develop music understanding systems by means of understanding
music understanding''.}
That is the central tenet of this manifesto.
An important consequence of adopting this is to put the listener into the center
of our focus, but in the role of an \textit{active listener} rather than the general
user-as-consumer with `listening habits' and `entertainment needs'
as discussed in \cite{schedl:etal:jiis2013}.

Before proceeding, a qualification is in order.
This text focuses exclusively on \textit{Western (tonal) music}.
Likewise, when speaking about listeners, listeners' expectations, etc.,
I will mean persons who have grown up with Western music
and thus have a shared (if implicit) understanding of music and stylistic norms.
Of course, there are many other important musical traditions
in the world, and the MIR community is increasingly addressing these
as well (e.g., in the CompMusic project \cite{serra:ismir2011}).
But looking at different musical cultures is beyond the scope of the present paper.

\section{Six Theses About Music}
\label{sec:theses}

I would like to structure my critique of the current state of the art by proposing
-- as is expected of a \textit{manifesto} --
six \textit{theses} about music and its effect on listeners.
None of these is probably controversial in itself -- in fact, the theses are
rather trivial truisms --, but each of them
points to a particular aspect that has, I believe, not received enough
attention in the MIR community. Each of the theses, if taken seriously,
has implications for future research, which I will try to work out in the following.

\subsection*{I.\ \ Music is a temporal construct / process}

Music as played and heard is a process that unfolds in time.
Even when viewed as a static object (represented, e.g., by the printed score),
a musical composition is organised along an abstract time line, where the ordering
of musical events, and their placement on the abstract time grid, is essential.
Clearly, then -- and that has been noted by many authors -- the (still) predominant
\textit{bag-of-frames (BOF)} approach to many music classification tasks is inadequate.
It is inadequate even as a summarising model of the perceived \textit{sound} of a piece
\cite{aucouturier:pachet:2004}, and definitely as a model of a piece
as heard by listeners, or intended by the composers and performers.
For instance, its limitations for emotion recognition have been convincingly
demonstrated in \cite{huq:etal:jnmr2010}.

The obvious alternative -- \textit{temporal models}, or at the very least, features that
incorporate some contextual information -- has been examined by several authors,
with mixed success. For instance, \cite{flexer:etal:dafx2005} showed that using
Hidden Markov Models (HMMs)
for timbre modelling increases the likelihood of the models
(i.e., the fit on the data), but does not improve similarity-based genre classification. 
Other authors \cite{madsen:etal:ismir2014,vaizman:etal:ismir2011} demonstrate some
improvement in emotion detection from audio, using temporal models.
In all these cases, however, very simple, low-level audio features (MFCCs)
were used as a basis. But as \cite{aucouturier:pachet:2004} already concluded
in 2004: summarising statistics over low-level sound features will not permit
our systems to surpass a certain level of performance 
(the much-cited \textit{`glass ceiling'}), and this statement still seems valid.

The conclusion then must be that -- temporal models or not -- 
frame-based audio features are fundamentally the \textit{wrong representation level}.
Simple intuition tells us that much of what we consciously perceive or expect in
a piece is at the level of \textit{events} -- notes, chords, etc. --
not short windows of sound textures.      
Consider the Beethoven \textit{Les Adieux} example from above.
The three opening chords of the first movement are clearly heard, by any listener, as three
distinct events, and the listener is suprised at the onset of the third chord,
not ``somewhere between audio frames 710 and 725, where the
distribution of chroma vectors changes''.
This is even more important because
most pianists tend to slightly delay the third chord,
% (see also thesis VI below),
thus heightening the level of
expectation in the listener and, concomitantly, the level of surprise as the third chord
is not the expected harmony.\footnote{
For the uninitiated reader: the third chord is a surprising C minor instead of the Eb major
that listeners will (consciously or unconsciously) expect. The resulting chord progression,
notated as I-V-vi, is appropriately called a \textit{deceptive cadence} in music theory.
It is not too abundant in classical music, and extremely rare at the very beginning of
pieces -- and thus all the more unexpected and surprising here.}
It is only by perceiving the passage as made up of discrete events that the delay
of the third event can be registered at all, and have the effect that it does.
Likewise, a prediction of whether a song would be perceived as `boring' or `interesting'
(which, among other things, will have to rely on some estimate of the
musical complexity of the song \cite{russell:1982}), will probably not be possible
on the basis of variance measures at the short-term feature level.

My hope, then, is to see more research in the future on temporal modelling of
music \textit{at the level of musically meaningful (partly discrete) events
and patterns}.\footnote{
This may also help avoid the \textit{Clever Hans} effect
recently identified in various MIR systems \cite{sturm:ieee2014}, which
is clearly related to these systems focusing on features at
musically irrelevant levels.}
The fact that our current technologies for tasks like
audio source separation, onset detection, note transcription, chord identification,
or melodic/motivic pattern discovery from audio are still notoriously unreliable and
brittle (though even the latter has recently been shown to be at least partly
feasible \cite{collins:etal:aes2014}), makes this particularly challenging.

\subsection*{II.\ \ Music is fundamentally non-Markovian}

To my knowledge, almost all temporal models used in MIR are either of
a Markovian kind, assuming a strictly limited range of dependency of the musical
present on the musical past (as in HMMs, but also, for instance, in
auto-regressive features or `dynamic textures' \cite{barrington:etal:ieee2010}),
or have a kind of decaying memory (as in simple Recurrent Neural Networks (RNNs)).
On the other hand, it seems clear that music is of a fundamentally \textit{non-Markovian}
nature. Music is full of long-term dependencies:
most pieces start and end in the same key, even if they modulate to other tonalities in between;
themes return at regular or irregular intervals, after some intermittent material;
harmonic rhythm (the rate of change of harmonies) is similar in similar passages;
and so on. Moreover, musical units (segments, phrases, etc.) tend to have
certain lengths in terms of number of bars (often a power of two), and listeners are used
to expecting the return of a refrain after a certain number of bars.
This means that we need the ability to \textit{count}, which low-order Markov models
are incapable of.\footnote{
Some of these problems are addressed in recent work on \textit{Markov Constraint}
models, such as \cite{roy:pachet:aaai2013}, which proposes a solution for the counting
problem in musical meter  and, more recently, \cite{papadopoulos:etal:cp2015}, which
presents a method for sampling Markov sequences that satisfy some regular constraints
(represented by an automaton).}

What is needed, first of all, is to broadly acknowledge the non-Markovian
nature of music and be critically aware of the fundamental limitations of
HMMs and similar models in describing music. I do not always see that in the MIR literature.
Second, we need more research on complex temporal models with variable degrees
of memory. An example of recent work attempting to create a model that accounts
for temporal dependencies in polyphonic music is \cite{boulanger:etal:icml2012},
which employed a complex, hybrid network made up of Restricted Boltzmann Machines
(RBMs) and RNNs. However, evidence
that the network does capture temporal dependencies is only indirect (via likelihoods
and prediction accuracies), and in the end the authors had to conclude that long-term
structure seems still out of the model's reach.
Perhaps more promising are recent advances in RNNs with Long-Short-Term Memory (LSTM) units
\cite{hochreiter:schmidhuber:1997}. Actually, \cite{eck:schmidhuber:2002} showed
already in 2002 that LSTMs are capable of learning longer-term dependencies and structure
in music.
More recent work in MIR using LSTM networks has mostly focused on low- to mid-level tasks
such as onset detection \cite{eyben:etal:ismir2010},
note transcription \cite{boeck:schedl:icassp2012},
or chord identification \cite{sigtia:etal:ismir2015},
where rather local context is sufficient.
A lot more work is needed on models that can predict
musical events and patterns over longer timespans.

Complementing this, it will be important to invest more research efforts into
learning structural \textit{abstractions}, over which temporal dependencies can then
be modelled. Much of tonal music has a multi-level, often hierarchical organisation,
with higher-level building blocks made up of smaller patterns (e.g., the ubiquitous
ii-V-I chord progressions in Jazz). At a high level, one could then get by
with low-order Markov dependencies. Hybrid architectures that can learn
multi-level abstractions and temporal relations simultaneously (as,
allegedly, \textit{Hierarchical Temporal Memory (HTM)} \cite{hawkins:george:2006} can),
would be particularly attractive.

\subsection*{III.\ \ Music is perceived by human listeners}

In much of current MIR research, a recording is taken directly as a
representation of a piece of music, from which computers then extract
patterns such as beat, segment structure, etc.
This pragmatic approach may be sufficient for practical applications such as music
synchronisation or indexing, but when our goal is to predict more refined human
categorisations (such as, e.g., emotions or interestingness), we need to remember
that the ultimate place of music is in the \textit{human mind} \cite{wiggins:etal:2010}:
what we hear and how we respond to music is a product of an active
\textit{process of perception}, and only by understanding that process will we
ultimately be able to predict some of music's effects.\footnote{
Actually, a full account of music perception would even go beyond the `mind', acknowledging
that the body of the listener and social interactions also play an important role.
Current theories on \textit{embodied and social cognition} (e.g., \cite{leman:2008})
are highly relevant, but to keep the presentation focused
(and, admittedly, for a lack of conrete ideas on how to adequately address this
aspect),
I will leave these out of the present discussion.
}

Human musical memory, and our conceptualisation of a piece of music,
critically rely on abstraction and grouping. Humans are exceptionally good
at segmenting the stream of musical events into meaningful units,
on-line, while listening \cite{deutsch:2013}.
In trying to explain this, music psychologists often appeal to the `laws' of
\textit{Gestalt psychology} \cite{wertheimer:1923}.
Lerdahl \& Jackendoff's [1983]
% \citet*{lerdahl:jackendoff:1983}
highly influential \textit{Generative Theory of Tonal Music}
derives various grouping rules from such Gestalt
principles, in different structural dimensions
(grouping, meter, hierarchical pitch abstraction, tension/release) with intuitively
convincing structural predictions on selected classical music examples
(though the Gestalt approach to music segmentation has also been
challenged \cite{bod:jnmr2002}).
Implicitly, some of these Gestalt concepts also play a role in current music
segmentation algorithms \cite{paulus:etal:ismir2010}
(e.g., similarity of recurring segments as a grouping criterion, or local changes
in some feature dimensions as indicators of boundaries), but the
hierarchical abstraction (`time-span reduction') and tension-release 
(`prolongational reduction') models, which are particularly interesting from
a music perception point of view, have not found their way into the world of practical MIR.
One reason is that the rules as given are not free of ambiguities, contradictions, and
cyclic dependencies, which so far has prevented researchers from fully implementing
the theory even at the level of symbolic scores \cite{hamanaka:etal:jnmr2006}.
I do believe it would be worthwhile to dig further into this
(perhaps via probabilistic modeling, to address various inconsistencies).
The ultimate challenge will be to apply similar principles to musical grouping
at the audio level, and combine this with the best of current MIR audio segmentation
algorithms.

A second aspect that is vital to our perception and appreciation of music
is the \textit{dynamics} of the listening process, the permanent ebb and flow of
tension and release; of anticipation and realisation; of expectations emerging
about what is to come next (and when), and their confirmation or denial.
Authors like \cite{meyer:1956,narmour:1992,huron:2006} argue that this is a major
source of the aesthetic and emotional effect of music, and the reason why we may be
enthralled by a piece, or lose interest.
That there is a correlation between the predictability of certain musical
features, and the emotional response reported by listeners, has also been
shown in \cite{dubnov:etal:2006}.
In modelling this, it seems natural to take an information-theoretic approach,
using notions like conditional entropy and information content to quantify
the listener's uncertainty about what is to come next, and her
surprise, or lack thereof, at what really comes next.
Such models of musical expectancy have been advocated by several
researchers in recent years \cite{abdallah:plumbley:2008,pearce:wiggins:2012,temperley:2007}.
While the principal ideas are extremely elegant and appealing, there
are severe problems in applying them to non-trivial kinds of music --
in particular, how to model joint probability distributions over huge spaces
of musical events.
Current experimental models evade this by making strong Markov assumptions
% \cite{abdallah:plumbley:2008}
and restricting the experiments to
strictly monophonic \cite{pearce:wiggins:2012} -- even isochronous
\cite{abdallah:plumbley:2008} -- music. They are thus useful as theoretical
models of musical learning and expectancy, but not yet for practical applications.
Again, my conviction is that the (only) approach to making this tractable for complex
music is by \textit{abstraction}, i.e., modelling music in terms of higher-level
patterns.

An interesting aspect of these information-theoretic models, as indicated by first
experimental results, is that they can also
help in predicting perceived grouping and segment boundaries
\cite{pearce:etal:perception2010}.
I believe it would be extremely important to carry this kind of work further,
towards more complex and realistic musical scenarios.

\subsection*{IV.\ \ Music perception and appreciation are learned}

The question of which fundamental mechanisms -- if any -- of music perception
are innate, and which ones are learned, is interesting but beyond the scope of this paper.
Arguably, all of the higher-level patterns and the `meanings' of music are learned,
to a large extent simply through exposure \cite{patel:2008}.
That gives us hope to also make substantial progress in machine
understanding of music via massive unsupervised learning.
The potential of statistical learning for explaining the emergence of musical
expectation has been demonstrated in \cite{pearce:etal:2010,pearce:wiggins:2012}.
Ongoing advances in the field of \textit{deep learning architectures} \cite{bengio:2009}
now promise to provide a general basis for learning
features and representations directly from raw data \cite{humphrey:etal:jiis2013}.
Deep learning models
have shown promise in several MIR tasks, from speech and music detection
\cite{schlueter:sonnleitner:dafx2012},
to audio segmentation \cite{ullrich:etal:ismir2014},
but also predicting performers' expressive dynamics from scores in classical
piano music via score features learned in this way \cite{grachten:krebs:ieee2014}
(to cite just some of our own work).
Particularly attractive are unsupervised learning scenarios, which promise to
make it possible to exploit
large musical datasets without the need for expensive manual annotation.

This is now the time for the MIR community to embark on massive feature / representation learning
endeavours -- much like the current trend in image analysis, which starts to produce
quite spectacular results (e.g. \cite{he:etal:arxiv2015}).
Given the computational and data-related demands, the MIR community should join forces
and pool its resources, efforts, and learned models (in cases where the
training data itself cannot be shared) -- and indeed, it has already begun to do so
(see \cite{porter:etal:ismir2015,weyde:etal:ismir2015} for two recent initiatives).
% \cite{krizhevsky:etal:nips2012}.
Also inspiring is recent work on networks that learn to verbally describe the content
of images \cite{vinyals:etal:arxiv2015}.
% Large-scale 
Music search engines that support description-based search without
the need for manual annotation would be extremely useful not only in the general
consumer music market, but also in specialised domains such as `production music',
where customers search for music with very specific properties.

Two aspects of music should again be kept in mind:
the \textit{multi-level} structure of music that ranges from timbre and sound through notes,
chords, rhythmic patterns, harmonic patterns (e.g., cadences), melodic motifs, themes, sections, etc.;
and the fact that different music-parametric dimensions (melody, harmony, rhythm)
\textit{interact} in complex ways. Learning useful and musically effective representations will
benefit from a careful design of learning architectures, guided, wherever possible,
by thoughtful analysis of the nature of music.

Long-term style learning would lead to building blocks like
typical cadences, typical harmonic progressions, melodic clich\'es, accompaniment
patterns, and the like.
But interactive real-time systems (such as an automatic music accompanist)
will also need learning at a different
time scale: \textit{short-term, on-line, intra-piece learning}, to induce some
of the characteristics of the currently playing song.
This is how we develop, during listening, very specific expectations about how a song
is going to continue, and when to expect certain things (like the next chord change, or
the return of the refrain).
There has been relatively little research on this in the MIR community that I am aware of;
most relevant is work on machine improvisation (e.g.,
\cite{assayag:dubnov:2004,nika:chemillier:2012,pachet:jnmr2003}),
or on learning to anticipate the timing of events in expressive
performance models \cite{raphael:icml2010,arzt:widmer:smc2010}.

A big open problem is how to \textit{integrate} long- and short-term learning.
It is not at all obvious according to what principles that should be done.
The \textit{IDyOM} model of \cite{pearce:wiggins:2012}, which
in its current form is a learning-based model of melodic prediction
(at a symbolic level), simply predicts a probability distribution over the next pitch
that is a weighted sum of the predictions of a long-term and a short-term learning model
(weighted by the respective Shannon entropies).
For more complex learning and prediction tasks, this will become more difficult.
Unfortunately, there is precious little that music psychology and cognition research
can tell us about this, in concrete terms. We may also see this as an opportunity:
new computational or information-theoretic models we may come up with
might serve as a source of inspiration to the music psychology world.

\subsection*{V.\ \ Music is (usually) performed}

In almost all kinds of music, musical compositions are performed (translated
into sound) by human musicians, and the details of the performance
contribute much to the character of the music, and how it affects
listeners. Expressive performance serves several functions \cite{palmer:1997},
most importantly, to clarify the musical structure of a piece to the listener,
and to highlight and communicate expressive and affective qualities of the music
(see also item VI below).

Surprisingly, the aspect of performance
has not seen too much attention in the MIR literature so far.
% For instance, in the recent three ISMIR conferences (2013-2015), there is maybe
% a dozen papers that explicitly focus on performance aspects.
That is a pity, as expressive performance can contribute much to the effect of music,
and to qualities that one may like or dislike in a recording.
Sophisticated music recommenders and other services should be aware of that.
Moreover, interactive music systems, such as the automatic accompaniment system
mentioned in Section \ref{sec:need} above, will need the ability to recognise and
emulate different performance and expression styles, in real time
(in addition to being able to anticipate expressive timing, as in \cite{raphael:icml2010}).

Many of the features we need to extract from a recording in order to account for
performance aspects, are rather different from the audio features mainly in use in
MIR. For instance, the `flowing' character of Jo\~ao Gilberto's renditions of
Bossa Nova songs -- in contrast to Frank Sinatra's singing or Stan Getz's
saxophone playing --
is due to the fact that he sings \textit{against} or \textit{`above'} the steady
beat of the song, taking incredible and at the same time extremely natural-sounding
liberties (a characteristic of Gilberto's art of Bossa Nova singing).
% \footnote{Listen to them (Jobim and Sinatra) side by side in a duet
% \url{https://www.youtube.com/watch?v=nC8in1DeWdQ}. Girl from Ipanema starts at 4:40.}
Current beat tracking algorithms (e.g., \cite{krebs:etal:ismir2015}) would readily
recognise the steady 4/4 beat; but characterising this floating on top of the beat
requires additional, in a sense `orthogonal' features.

Depending on the instrument, there is a large variety of parameters that
performers can control and shape, from tempo, timing, loudness, articulation 
to complex continuous aspects such as intonation, vibrato, timbral control of the singing
voice\footnote{To get an impression of the richness of expressive
possibilities in vocal art, listen to any recording of, say, Sarah Vaughan
or Abbey Lincoln.}, etc. We currently do not even have features that can quantify the degree of
`staccatoness' vs. `legatoness', much less methods for exactly measuring
micro-timing in chords, the sound balance between individual voices in a polyphonic piece,
or qualities of singing.
It would be desirable, at some point, to have such performance-related features included
as a standard part of MIR feature extraction toolkits, alongside the current feature
sets describing timbre (e.g., spectral centroid, MFCCs), rhythm (e.g., beat histograms),
and the like.

Recognising and characterising performance-related aspects in music is one problem.
Another is to build \textit{predictive models} that can % \textit{anticipate} and
\textit{produce} performances with certain musical qualities.\footnote{
Again, the automatic accompanist is a use case for this; quite another one would
be systems that adapt the expressive character of music to dynamically changing
situations, e.g., in video games or interactive movies.}
In the context of classical music, there has been quite some research on
computational models of expressive music performance, as summarised in
a 2004 survey \cite{widmer:goebl:jnmr2004} -- and the state of the art has not really
improved a lot since then (though YQX has successfully performed Chopin in
a RENCON computer piano performance contest \cite{widmer:etal:aimag2009}).
The group of researchers working on computational performance analysis and modelling
has traditionally been quite small. Bringing the power of the full MIR community
(with its interests that extend way beyond the narrow world of classical music)
to bear on this class of problems would be extremely promising.

\subsection*{VI.\ \ Music is expressive and affects us}

Approaches to automatic music recommendation have been rather superficial so far,
evading the issue of what listeners actually hear, and \textit{why} they might like a song.
Typical music recommenders rely on indirect information such as
timbral and rhythmic similarity between songs, expert-curated or web-crawled meta-data, 
user-provided tags, collaborative filtering, and/or features characterising the
geographical and activity context of users \cite{song:etal:cmmr2012}.

But music is more than that. It moves us; it affects us; a song may cause us to
feel elated or sentimental; we may be touched by the mourning, solemn character of
a funeral march.\footnote{
Be aware of the difference between the \textit{arousal} of emotions, and the
`mere' \textit{expression} or \textit{communication} of emotions \cite{gabrielsson:2002}
-- a distinction that is not always clearly made in the MIR % emotion recognition
literature. In either case, however, the ability of music to express emotions and moods,
or to \textit{incite} [D.~Cope, personal communication]
listeners to construct musical and affective meaning, considerably adds to its
importance and power as an art form.},\footnote{
A lucid (and short) discussion of different philosophical views on this
can be found in \cite{london:2000}.}
I strongly believe that MIR systems should be aware of this dimension, at least
to the extent that they can find music for us that
really has the potential to satisfy our musical and affective needs.
The recent increase in emotion recognition research
\cite{kim:etal:ismir2010,yang:chen:acmtist2012} shows
that the MIR community acknowledges the importance of that dimension.

However, I contend that the set of qualities that music can express is
much broader than `just' emotions or moods, and thus broader
than the dimensions usually targeted in emotion detection.
\cite{juslin:2013} distinguishes three levels of `coding'
of expressive messages in music, and through that, implicitly, different
classes of expressive content:
while basic emotions are communicated via rather direct cues like
loudness, tempo, vocal qualities in singing -- and are thus perhaps most directly accessible
to  MIR systems via audio features --, more complex and abstract qualities arise
from the structure of the music itself, its implications of melodic/harmonic/rhythmic tension,
release, realisation or denial (see above). Juslin calls this \textit{intrinsic coding}
and suggests that such factors, ``[b]y contributing dynamically shifting levels of
tension, arousal and stability, [...] may help to express more complex, time-dependent
emotions, such as relief and hope''
-- to which one might add such qualities as uncertainty, determination, humour,
but also power and physical motion and other things that I would not subsume
under categories like emotions or moods.
Juslin's third level of meaning assignment -- \textit{associative coding} -- relates to
expressive meanings that are purely conventional and socially learned,
such as that a song that is presented to us as a national anthem
conjures up images of patriotic pride or nationalism (as the case may be).
Such meanings are not necessarily related to any specific properties of the music
itself.

From the above follow several research challenges.
First of all, there is  a need for broad empirical investigations on what kinds of
expressive qualities humans can (relatively) reliably and consistently recognise
in music. Second, we need to categorise these, and define appropriate vocabularies or description
frameworks. I believe that the popular categorisation models for emotions -- e.g.,
valence-arousal space \cite{thayer:1989}, `circumplex model' \cite{russell:1980},
Geneva Emotional Music Scale (GEMS) \cite{zentner:etal:2008} -- cannot capture all the
expressive qualities that music can convey.
The set of categorical `mood adjectives' used in the MIREX Mood Classification Task for
popular music\footnote{
\url{http://www.music-ir.org/mirex/wiki/2013:Audio_K-POP_Mood_Classification}}
contains a number of interesting concepts that actually go beyond moods proper
(for instance, I would claim that `rowdy', `whimsical', or `literate' are neither
emotions nor moods). But again, I fail to see any systematic evidence
that this set covers all the qualities we can and want to distinguish.
% (also in classical music).

In designing algorithms that can recognise and classify expressive dimensions,
different sources of expressivity must be considered.
Simple \textit{`surface properties'} like tempo, dynamics, timbre and chosen
instruments, mode (major/minor), seem to most directly communicate
(and partly even induce) basic emotions \cite{juslin:2013}. Then there is
the \textit{structure} of the composition itself, with its ups and downs, games of tension and release,
and more or less dramatic twists and turns, which are more challenging to capture
in terms of features (see above).
\textit{Culturally defined} meanings whose source is outside the music itself
will only be accessible or inferrable to MIR systems from extra-musical sources
-- particularly, the Web  (e.g., \cite{knees:schedl:acmtmcca2013}).
A final aspect that has been largely ignored so far in the MIR world is, again, \textit{performance}.
Especially in classical music, the specific way in which performers play a piece
has a tremendous influence on the perceived character of the resulting music
(see the little \textit{Con Espressione Game} in the next section) -- but this
also goes for Jo\~ao Gilberto's or Sarah Vaughan's singing, or any other performed
music (including, yes, \textit{Kraftwerk}).
The \textit{Con Espressione} project will place a special focus on performance as a
source of expressivity, but all of the above description levels will be needed,
as performance cannot be seen independently of the piece itself and its structure
\cite{gabrielsson:lindstroem:2010}.

\section{The \textit{Con Espressione} Project}
\label{sec:conespressione}

\textit{Con Espressione} is a five year research endeavour (2016-2020)
funded by the European Research Council (ERC)
% in the form of an ERC Avanced Grant
(\url{http://www.cp.jku.at/research/projects/ConEspressione}).
Its goal is to lay the foundations for a new generation of music
systems that are aware of, or can recognise and characterise, \textit{expressive}
aspects of music. The primary focus (owing to my research team's extensive experience
and prior work) will be on classical music and expressivity as communicated via expressive
\textit{performance}.
In approaching this, we will have to address some of the challenges discussed above.
Specifically, we will
\begin{itemize}
 \item investigate description frameworks for characterising and categorising
       (intended and perceived) expressive dimensions;
 \item advance research on extracting performance parameters (beyond timing
       and dynamics) from audio recordings and live performances;
 \item work on computational models of structure perception in music
       (at score and audio levels), combining recent MIR advances with information-theoretic
       approaches, unsupervised learning, and knowledge from musicology;
 \item investigate the relation between musical structure, expressive performance,
       and the communication of expressive characters;
 \item learn discriminative models that recognise intended expressive messages
       in performances;
 \item learn predictive models that generate or modify performances to express certain
       intended qualities. 
\end{itemize}

All this will involve large-scale machine learning, using large curated corpora
currently in preparation.
One of the demonstrators we hope to deliver by the end of the project is
the \textit{Compassionate Music Companion}, an interactive system that
plays along with human musicians in a musically sympathetic way, recognising, anticipating,
and adapting to the musicians' expressive way of playing,
and providing musical interaction at a gratifying level.

\subsection*{The Con Espressione Game}

I would like to take the opportunity to invite the reader
% (as I will invite all audiences in Con-Espressione-related presentations)
to the \textit{Con Espressione Game}:
% (see Fig.~\ref{fig:mozart:game}):
the link {\tt\url{bird.cp.jku.at/con_espressione_game}} will take you
to a page that asks you to listen to excerpts from five Mozart sonata renditions
by different pianists, and to enter words which, to you, best describe the perceived
character of the recordings. This is a first small test to see to what extent (if at all)
there is consensus in the perception of expressive qualities in classical
piano performances. The collected responses will be analysed and the results announced to
the MIR community in due course.

\section{Conclusion}

This little manifesto has reminded the reader of a few simple truths
about music (rather pompously called `theses' here), and what they
might imply in terms of research challenges for our field:

\begin{itemize}
  \item [ ] \ \ i.\ \  Music is a temporal construct / process
  \item [ ]  \ ii.\ \  Music is fundamentally non-Markovian
  \item [ ]   iii.\ \  Music is perceived by human listeners
  \item [ ]  \ iv.\ \  Music perception and appreciation are learned
  \item [ ] \ \ v.\ \  Music is (usually) performed
  \item [ ]  \ vi.\ \  Music is expressive and affects us
\end{itemize}

The ERC project \textit{Con Espressione} will expressly try to address some of the
questions that follow from these principles, focusing on the expressivity
of music and music performance.
My hope is that this manifesto will stimulate other research teams to join in
this effort, so that future MIR systems will understand a bit more
of the \textit{essence} of music, and will be able to provide services at a
new level of musical quality.

% Acknowledgments
\begin{acks}
The author gratefully acknowledges generous long-term financial support for this research
by the Austrian Science Fund FWF (Wittgenstein Award 2009, project number Z159)
and the European Research Council (ERC Advanced Grant, project number
ERC-2014-AdG 670035).
Thanks to David Cope, Fran\c{c}ois Pachet, Perfecto Herrera,
Arthur Flexer, Werner Goebl, Thomas Grill, Andr\'e Holzapfel,
Rainer Kelz, Peter Knees, Florian Krebs,
Markus Schedl, Jan Schl\"uter, and Reinhard Sonnleitner,
as well as two anonymous reviewers,
for helpful comments and hints, with apologies for not having been able
to incorporate all the (very appropriate and useful) comments in a
short article like this.
\end{acks}

% Bibliography
\bibliographystyle{ACM-Reference-Format-Journals}
\bibliography{manifesto}
                             % Sample .bib file with references that match those in
                             % the 'Specifications Document (V1.5)' as well containing
                             % 'legacy' bibs and bibs with 'alternate codings'.
                             % Gerry Murray - March 2012

% History dates
\received{October 2015}{January 2016}{February 2016}

\end{document}